\documentclass[pdflatex,sn-nature]{sn-jnl}%

\usepackage{dsrm_local}%
\usepackage{graphicx}%
\usepackage{multirow}%
\usepackage{amsmath,amssymb,amsfonts}%
\usepackage{amsthm}%
\usepackage{mathrsfs}%
\usepackage{xcolor}%
\usepackage{textcomp}%
\usepackage{manyfoot}%
\usepackage{booktabs}%
\usepackage{algorithm}%
\usepackage{algorithmicx}%
\usepackage{listings}%

\usepackage[frozencache=true,cachedir=minted]{minted}
\usepackage[caption=false,font=footnotesize]{subfig}%

\newcounter{objective}[section]
\newenvironment{objective}[1][]{\refstepcounter{objective}\par\noindent\textbf{O\theobjective~-~#1}: \rmfamily}

\makeatletter
\let\myorg@bibitem\bibitem
\def\bibitem#1#2\par{%
  \@ifundefined{bibitem@#1}{%
    \myorg@bibitem{#1}#2\par
  }{%
    \begingroup
      \color{\csname bibitem@#1\endcsname}%
      \myorg@bibitem{#1}#2\par
    \endgroup
  }%
}

\lstnewenvironment{tabularlstlisting}[1][]
 {%
  \lstset{aboveskip=-1.3ex,belowskip=-3.5ex,#1}%
 }
 {}

\begin{document}

\title[Article Title]{Toolchain for Faster Iterations in Quantum Software Development}

\author*[1]{\fnm{Otso} \sur{Kinanen}}\email{otso.j.r.kinanen@jyu.fi}

\author*[1]{\fnm{Andrés D.} \sur{Muñoz-Moller}}\email{andres.d.munozmoller@jyu.fi}

\author*[1]{\fnm{Vlad} \sur{Stirbu}}\email{vlad.a.stirbu@jyu.fi}

\author*[2]{\fnm{Juan M.} \sur{Murillo}}\email{juanmamu@unex.es}

\author*[1]{\fnm{Tommi} \sur{Mikkonen}}\email{tommi.j.mikkonen@jyu.fi}

\affil*[1]{\orgname{University of Jyväskylä}, \orgaddress{\city{Jyväskylä}, \country{Finland}}}

\affil*[2]{\orgname{Universidad de Extremadura}, \orgaddress{\city{Cáceres}, \country{Spain}}}

\abstract{Quantum computing proposes a revolutionary paradigm that can radically transform numerous scientific and industrial application domains. To realize this promise, these new capabilities need software solutions that are able to effectively harness its power. However, developers may face significant challenges when developing and executing quantum software due to the limited availability of quantum computer hardware, high computational demands of simulating quantum computers on classical systems, and complicated technology stack to enable currently available accelerators into development environments. These limitations make it difficult for the developer to create an efficient workflow for quantum software development. In this paper, we investigate the potential of using remote computational capabilities in an efficient manner to improve the workflow of quantum software developers, by lowering the barrier of moving between local execution and computationally more efficient remote hardware and offering speedup in execution with simulator surroundings. The goal is to allow the development of more complex circuits and to support an iterative software development approach. In our experiment, with the solution presented in this paper, we have obtained up to 5 times faster circuit execution runtime, and enabled qubit ranges from 21 to 29 qubits with a simple plug-and-play kernel for the Jupyter notebook.} %

\keywords{Quantum software, software development, developer experience, Kubernetes, Jupyter notebooks}

\maketitle

\section{Introduction}

Quantum computing holds great promise as a revolutionary technology that can transform various scientific and industry fields. By harnessing the principles of quantum mechanics, quantum computers can perform complex calculations and solve problems that are currently intractable for classical computers. This promises breakthroughs in areas such as cryptography, optimization, drug discovery, materials science, or machine learning. %

Although quantum advantage has been declared in experiments where quantum computing hardware has shown to provide a significant computational advantage over classical alternatives in specific problems~\cite{sandersquantum}, we still have to work for the foreseeable future with Noisy Intermediate-Scale Quantum (NISQ) computers. These computers employ a hybrid computational model in which a classical computer controls a noisy quantum device build from a variety of qubits (e.g. superconducting~\cite{Clarke2008quperconducting}, trapped ions~\cite{Debnath2016trapped}, nuclear spins in silicon~\cite{Noiri2022} or photonic~\cite{Peruzzo2014}) that allows noisy initial state preparation, performing low fidelity quantum gates and noisy measurements. Even as NISQ devices are not capable of providing the quantum advantage promised by quantum algorithms~\cite{Chen2023}, they are an invaluable platform for research and experimentation.

Even with the steady advancements in Quantum Computing technology in terms of both, qubit counts and fault tolerance~\cite{preskill2023quantum, gill2022quantum}, and the increasing number of hardware vendors, the current NISQ computers still remain out of reach for constant use for many developers. This is due to hardware scarcity, vendor dependent development infrastructure and high operational costs of QPUs. Therefore, many quantum software developers rely on simulators running on classical computers to experiment with quantum software during the development process. While it is straightforward to start the development process locally on commonly used classical computing hardware, scaling up the development, necessitates running larger circuits on specialized more capable environments, with efficient simulators like graphical processing units (GPU) that may be found in high-end consumer products (e.g. mobile workstations) or in high-performance computing infrastructure (e.g. clusters of GPUs), and from there eventually forward to a Quantum computer. But to get the advantage provided by the GPUs a developer is currently required to have either deep technical knowledge to configure the software stack required for using the advanced GPU capabilities to simulate quantum circuits up to 31 qubits~\cite{faj2023benchmark}, or to have access to a supercomputer infrastructure enabling execution for circuits with up to 40 qubits~\cite{WILLSCH2022108411}.

Our approach to improving the workflow in quantum software development is 
building a toolchain with a goal to enhance the execution of quantum software routines and lower the barrier of moving between platforms. In this article, we focus on the usage of classical quantum computing simulators as part of that workflow and to enable the developer to move between execution platforms effortlessly while making advancement between development cycles and to have each code execution be as efficient as possible.
This is done by using remote computational resources efficiently and leveraging current best known simulators and GPUs in the execution. The solution we present to support these goals is packaged as an easy-to-use Jupyter kernel, in which the developers are not directly exposed to the complexities of operating the cluster where the quantum routines are executed. The solution allows an effortless transition from the local, to remote development execution environments. This will benefit the developer with noticeable time savings and added range as the executable circuits grow wider (in terms of qubit count) and/or deeper (in terms of operations applied). During the development process, the code under work is encouraged to be developed iteratively, therefore as the execution frequency increases, the role of each execution time adds up leading to a fragmented developer experience. Moving the computationally intensive and time consuming executions from the developer's local premises to the remote cluster smoothens their workflow, leading to the possibility to do faster and more frequent iterations during the development process.  

As the contribution of this research, we will present a \textit{practice} to scale up the execution platform from the local environment to the actual quantum computer, the \textit{tooling} required to support the proposed practice, and the \textit{results} obtained in our experiments with execution speedups in several quantum code benchmarks that demonstrate the improvements with regard to development time.

The rest of the paper is organized as follows. Section~\ref{sec:background} presents the background and motivation behind this work. Section~\ref{sec:objectives} introduces the methodology used to perform the study and the objectives of the solution. Section~\ref{sec:solution} describes the implementation of the solution. Section~\ref{sec:evaluation} describes the environment, in which we performed the evaluation of the solution, the impact on presented workflow models and addresses threats to validity. Section~\ref{sec:conclusion} concludes with some final remarks and presents the future work.

\section{Background and motivation}
\label{sec:background}
\subsection{Software development life cycle}
\begin{figure*}
    \centering
    \includegraphics[width=1\textwidth]{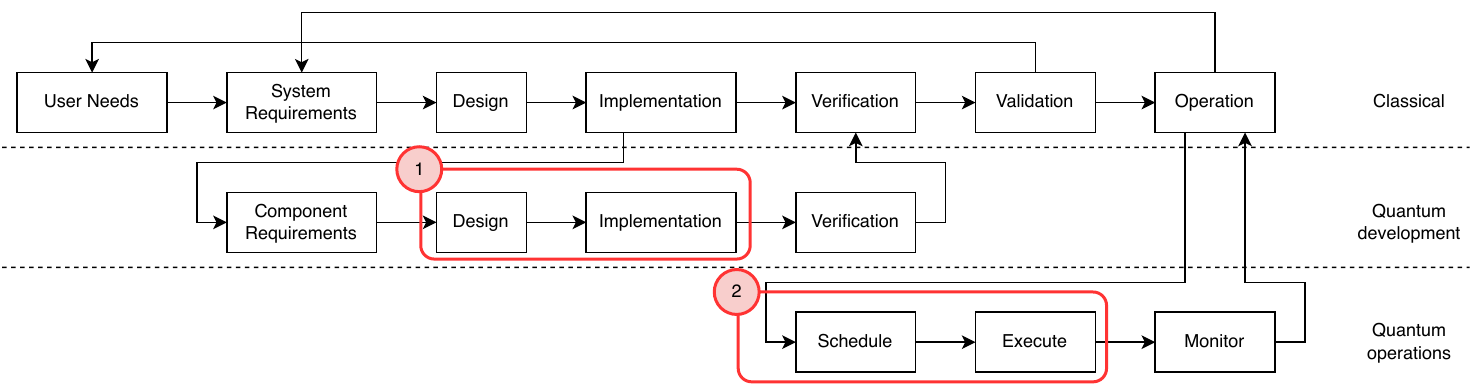}
    \caption{Quantum software development life cycle and areas where developers and operators interact with quantum hardware or simulators: (1) the design and development of quantum algorithms, and (2) scheduling and executing the computation on an available and capable quantum computer.}
    \label{fig:sdlc}
\end{figure*}

One of the foundational literature for quantum software engineering, Talavera manifesto suggests embracing the coexistence of quantum and classical computing when engineering a quantum system. \cite{piattinitalavera}. 
The system design should allow adapting quantum capabilities to classical software and into the development process. This leads to need of redefining the development life cycle commonly used in classical software development\cite{perezguidelines}.
The software development life cycle (SDLC) of hybrid classic-quantum applications consists of a multifaceted approach \cite{stirbu2023fullstack}, as depicted in Figure~\ref{fig:sdlc}. At the top level, the classical software development process starts by identifying user needs and deriving them into system requirements. These requirements are transformed into a design and implemented. The result is verified against the requirements and validated against user needs. Once the software system enters the operational phase, any detected anomalies are used to identify potential new system requirements if necessary. A dedicated track for quantum components is followed within the SDLC~\cite{sdlc}, specific to the implementation of quantum technology. The requirements for these components are converted into a design, which is subsequently implemented on classic computers, verified on simulators or real quantum hardware, and integrated into the larger software system. During the operational phase, the quantum software components are executed on actual quantum hardware. The scheduling ensures efficient utilization of the scarce quantum hardware resources, while monitoring capabilities enable the detection of anomalies throughout the operational stage.

As quantum computers are a limited resource, it is currently not practical to develop quantum software components directly on hardware. Instead, developers should use simulators that use commonly available and less expensive classical resources (e.g., CPUs and GPUs~\cite{GUTIERREZ2010283}) for the early stages of development and testing. %
When proceeding in the development process, developers may move to more sophisticated simulators that can simulate the noise of actual hardware. Only when the components are mature enough, the development should be continued on quantum processing units~(QPU), the actual hardware that will be used during the execution phase. However, as the implementation of quantum software stack trades off the visibility of the execution process for usability \cite{stirbu2024interplay}, developers have to experiment and iterate on devices and simulators to determine the actual behaviour of their programs. This approach ensures that the use of quantum resources is efficient and effective.

\subsection{Towards an iterative workflow for quantum development}

The SDLC for quantum and hybrid application includes a quantum development phase, which we examine closer as an internal process. The model suggests that for the quantum components, the developer should follow an inner development cycle, Quantum Circuit Lifecycle where the implementation of the  quantum software starts from the classical - quantum splitting, is followed by hardware-independent quantum circuit development, then by hardware selection and optimization, up to the execution on selected QPU, and finally analyzing the results \cite{sdlc}. 

In a model for quantum-classical system design proposed by Perez-Castillo et al.~\cite{perezguidelines}, Incremental Commitment Spiral Model (ICSM) for quantum, authors suggest an incremental and iterative approach. While these models have slightly different scope and point of view, they share similar core ideas for the quantum development phase. ICSM focuses on the broader perspective of system design and development processes. However, within the internal development cycles, it emphasizes an incremental workflow with continuous adjustments and evaluations in each iteration.

Next, focusing more closely on the inner cycle of the development process, Hardware Independent Implementation, where the quantum circuit is developed, tested and verified. As part of the models this is presented as one part of a cycle, but it is important to notice that, as well as in classical software development, the programming, testing and validating of the quantum code is yet an other iterative and incremental process itself. To make this part of the process as swift as possible on each iteration, every code execution within the cycle needs to be efficient, which may only be obtained with right selection of execution targets available from the development environment. 
While these models offer clear view on wider perspective of the development process, and suggest efficient process models, we have recognized a limitation of practical workflow, and related tooling support.

\subsection{Quantum simulation methods on classical hardware and scalability}

To simulate quantum computers and quantum circuits there are several known methods which have different characteristics and use cases. One approach to simulation is with state vector or density matrix simulation, where the full quantum states are simulated and maintained throughout the execution. These methods are resource intensive on memory, with memory demand growing exponentially with the qubit count, creating the bottleneck in simulating quantum circuits as the qubit count grows~\cite{xuherculeantaskclassicalsimulation}. For perspective, in experiments with a GPU cluster with 2048 NVIDIA A100 GPUs, accommodating 40 gigabytes of memory each, Willsch et al. reached limits at 42 qubits~\cite{WILLSCH2022108411}. 

Nonetheless, alternative methods are available to improve the efficiency of classical simulation for certain classes of quantum circuits. For example, quantum circuits containing few non-Clifford gates can be simulated efficiently up to a higher number of qubits using low-rank stabilizer decompositions \cite{Bravyi}. Similarly, efficient classical representation of quantum circuits using tensor network techniques, such as matrix product states (MPS) or Projected Entangled Pair States (PEPS) \cite{Ors}, allows one to simulate quantum circuits of an arbitrary number of qubits, as long as the MPS structure has a low bond dimension, that is, quantum circuits with a moderate degree of entanglement \cite{Patra}.

Regardless of the chosen simulation method, a developer can likely achieve performance improvements by executing the simulation in a more powerful environment—either by leveraging additional memory to simulate the full state or by utilizing efficient parallel computation on GPUs, e.g. with mentioned tensor network methods~\cite{tensornetworkmethodgpu}. That being said, later in this article, we concentrate on the more general case of state vector simulation.

\subsection{Computing at-scale paradigms}%

Cloud computing allows the development of scalable applications~\cite{cloudnativeapps}, which rely on computing resources like computing power, storage and databases that are accessed on a pay-per-use basis. Through the extensive use of application programming interfaces (APIs), teams formed of software developers and operators can scale these resources up and down in response to the users' needs. This entails designing applications as small, loosely coupled components that can be bundled with their dependencies into portable containers and deployed on the immutable infrastructure. Furthermore, integrated monitoring and logging offer valuable insights into performance, health, and behaviour, empowering a swift response to potential anomalies. 

Kubernetes is the industry-standard container orchestration platform for automating deployment, scaling, and management of containerized cloud-native applications \cite{luksakubernetes}. Developed as an open-source solution by Cloud Native Computing Foundation (CNCF)\footnote{\href{https://www.cncf.io/}{https://www.cncf.io/}}, together with the myriad of projects that offers supporting functionality, it allows users to deploy applications on the managed infrastructure of the major cloud providers (e.g., AWS EKS\footnote{\href{https://aws.amazon.com/eks/}{https://aws.amazon.com/eks/}}, Azure AKS\footnote{\href{https://azure.microsoft.com/en-us/products/kubernetes-service}{https://azure.microsoft.com/en-us/products/kubernetes-service}}, or GCP GKE\footnote{\href{https://cloud.google.com/kubernetes-engine}{https://cloud.google.com/kubernetes-engine}}), smaller or regional cloud providers, or on-prem -- using own infrastructure.

High-performance computing (HPC) relies on using supercomputers and parallel processing techniques to solve complex computational problems quickly and efficiently, in application domains that require massive computational power~\cite{sterling2017hpc}. HPC systems typically consist of multiple interconnected processors or nodes that work together to execute tasks in parallel, enabling large-scale simulations, data analysis, and scientific computations, leveraging the Open Message Passing Interface~(OpenMPI\footnote{\href{https://www.open-mpi.org}{https://www.open-mpi.org}}) compatible architectures.

Although cloud computing and HPC have distinct purposes -- on-demand access to computing resources online versus providing computing power for complex scientific and computational tasks -- they both face increasingly intense competition for the utilization of specialized accelerators like GPUs, a trend noticed by vendors that allow partitioning single GPU instances with techniques like Multi-instance GPU\footnote{\href{https://www.nvidia.com/en-us/technologies/multi-instance-gpu/}{https://www.nvidia.com/en-us/technologies/multi-instance-gpu/}}. Further, despite being operated in different ways -- public cloud providers or on-prem versus national laboratories, research institutions, and specialized HPC centres -- each approach has technical capabilities that are useful in the other domain.

For example, training machine learning models in Kubernetes with Kubeflow\footnote{\href{https://www.kubeflow.org}{https://www.kubeflow.org}} can take advantage of HPC-like resources via the MPI Operator\footnote{\href{https://github.com/kubeflow/mpi-operator}{https://github.com/kubeflow/mpi-operator}}. Similarly, the more sophisticated orchestration, monitoring capabilities, and integrations of the cloud-native computing have been identified as gaps by the HPC community~\cite{zhou2023hpcvsk8s}. The industry response was to establish the High Performance Software Foundation (HPSF)\footnote{\href{https://hpsf.io}{https://hpsf.io}} that aims to develop solutions that are aligned with Cloud Native Computing Foundation (CNCF)\footnote{\href{https://www.cncf.io}{https://www.cncf.io}}, the home of cloud-native development. We expect that in the long term, the technical implementations of the HPC and cloud-native computing to be much closer aligned than they are today.

Quantum computing enables the existing base of cloud-native and HPC applications to accelerate appropriate computational tasks. Two notable approaches for integrating the two software stacks are HPC-QC~\cite{hpc-qc-linking}, which uses the OpenMPI, and XACC~\cite{XACC} approach based on the OSGi\footnote{\href{https://www.osgi.org}{https://www.osgi.org}} architecture. Similarly, Qiskit's quantum-serverless~\cite{quantum-serverless} proposes a cloud-based approach for running hybrid classical-quantum programs. The proposed programming model, conforming to the RAY\footnote{\href{https://www.ray.io}{https://www.ray.io}} computing framework, makes it easy to scale Python workloads on a Kubernetes cluster in which the quantum execution environment is represented by a distributed Qiskit runtime that allows transparent access to multiple QPUs. Despite all these efforts, the integration of quantum computing into classical paradigms is fragmented. The EuroHPC aims to address this with the \textit{Universal Quantum Access}~\cite{eurohpc} development.

With the currently available tools, a quantum software developer has easy access to several separate tools and toolkits to begin their journey, but soon after, as the quantum circuits get more complicated to simulate the path gets complex. When going forward to more demanding quantum circuits, ranging from 20-30 qubits, the developer has to choose between building their own execution environment requiring investment in capable hardware, such as GPUs, and knowing how to build their own execution software infrastructure stack using these building blocks. The other possible way would be to use some of the external computation services with batch type of execution, e.g. HPC Clusters, or some with a slightly different approach using commercial cloud-based infrastructure with suitable hardware, demanding highly specific expertise to set up and to use.

\section{Methodology and objectives}
\label{sec:objectives}

\begin{figure*}[!t]
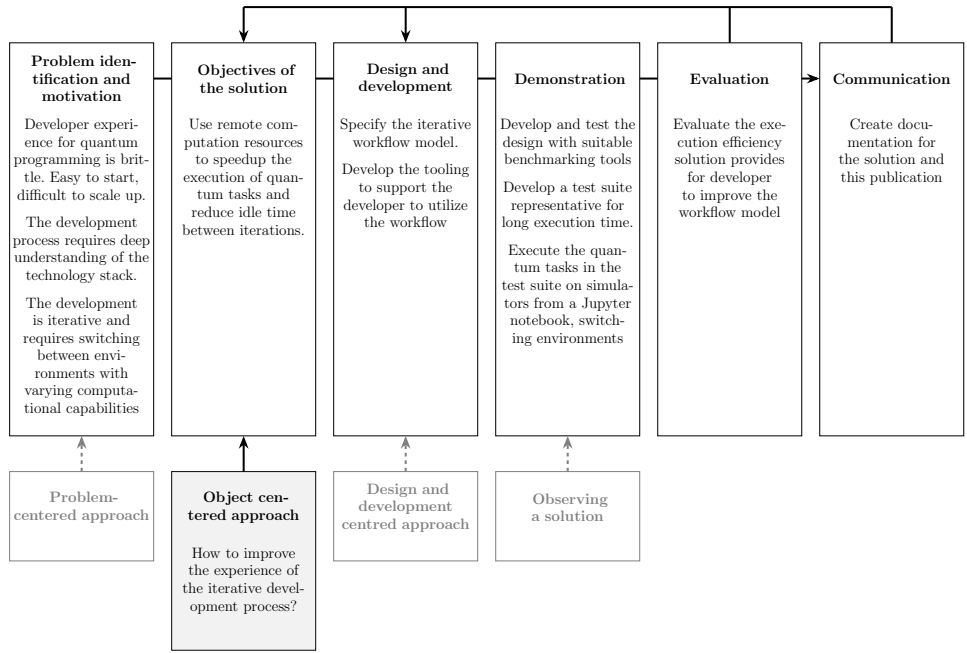

    \centering
    \resizebox{\textwidth}{!}{%
        \SDR{
            \ProblemIdentification{Developer experience for quantum programming is brittle. Easy to start, difficult to scale up.\\\vspace{8pt} The development process requires deep understanding of the technology stack.\\\vspace{8pt} The development is iterative and requires switching between environments with varying computational capabilities}
            \ObjectivesOfTheSolution[How to improve the experience of the iterative development process?]{Use remote computation resources to speedup the execution of quantum tasks and reduce idle time between iterations.}
            \DesignAndDevelopment{Specify the iterative workflow model.\\\vspace{8pt}Develop the tooling to support the developer to utilize the workflow}
            \Demonstration{Develop and test the design with suitable benchmarking tools\\\vspace{8pt}
            Develop a test suite representative for long execution time.\\\vspace{8pt}Execute the quantum tasks in the test suite on simulators from a Jupyter notebook, switching environments
            }
            \Evaluation{Evaluate the execution efficiency solution provides for developer to improve the workflow model}
            \Communication{Create documentation for the solution and this publication}
        }
    }
    \caption{Design science research methodology applied to the improve the quantum software development process.}
    \label{fig:dsr}
\end{figure*}

The study was developed using the objective-centric approach of the Design Science Research (DSR)~\cite{dsrm} methodology, a process depicted in Fig.~\ref{fig:dsr}. The starting point was to answer the research question: \textit{How to improve the experience of the iterative quantum software development process?} Based on the findings outlined in the background and motivation, the research question was further refined into a set of objectives:

\begin{objective}[Iterative workflow with simulators]
\label{obj:iteration}
Where in the earliest stages of quantum circuit development, like prototyping an algorithm, local execution may be efficient, as soon as the circuits grow wider and deeper during the development iterations, the processing power demands for simulators in use grows fast. To keep both the change of execution target between iterations fluent, and the execution runtimes short, the development environment needs to support the the iterative workflow. 
\end{objective}

\begin{objective}[Execution speedup]
\label{obj:speedup}
GPUs provide high efficiency to quantum code execution, but using them as a local resource is often not possible for an individual developer, and they need to be accessed through extra layers of infrastructure and network. Despite the overhead from using remote GPU, the developer still gets better experience, and access to simulators that can run circuits with a larger number of qubits and depth and simulate noise.
\end{objective}

\begin{objective}[Execution target selection]
\label{obj:target}
To balance the benefits of remote GPU execution - such as speedup - with the drawbacks of increased networking overhead on small-scale circuits, it is essential to make the process of selecting an execution target straightforward and efficient for developers.
\end{objective}

\label{obj:usability}

The design and development phase consisted of determining the configurations of a Kubernetes cluster that is able to effectively execute quantum computation tasks using CUDA-capable quantum simulators and developing a Jupyter kernel that allows sending the quantum computations (e.g. the content of notebook cells) to the cluster. The demonstration phase consisted of demonstrating the use of the Jupyter kernel for executing a quantum routine test suite on two Kubernetes clusters. During the evaluation phase, we have assessed the results collected during the demonstration phase. For the communication phase, we have prepared this report and published using an open-source model of the kernel code to GitHub.

\section{Tooling support: the Python kernel for Kubernetes}
\label{sec:solution}

\subsection{Quantum development toolkits and simulators}
Now we describe briefly the quantum development and execution tools that we have used as the technology stack to build our solution and the benchmarking introduced later in the paper. 

\begin{figure}[t]
    \centering
    \includegraphics[width=0.6\columnwidth]{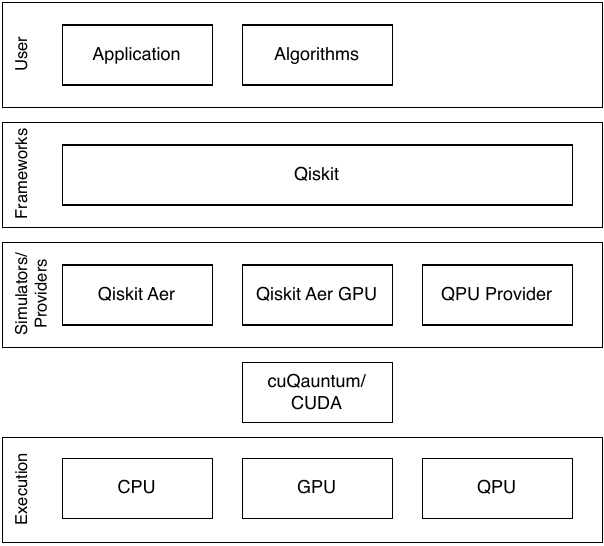}
    \caption{A layered view at the Qiskit software stack, where from top to bottom may be seen how User, Frameworks, Simulator back-ends, hardware drivers and the processor units build the execution stack for quantum algorithms or applications. }
    \label{fig:executionstack}
\end{figure}

Qiskit is a Python library and a quantum development toolkit designed to accommodate different types of quantum computers in the NISQ era. It allows algorithm designers to develop applications leveraging quantum computing, circuit designers to optimize circuits and explore its properties like error correction, verification and validation. Qiskit offers also tools to research and optimize gates, with precise control and the ability to explore noise, apply dynamical decoupling and perform optimized control theory. Qiskit is an open-source project and currently offers dozens of additional libraries, plugins, simulator backends, application packages for multiple domains such as machine learning, physics, chemistry and finance and other related projects available. In Qiskit there are also several transpiler plugins available for users to optimize and interact with the transpiling process\footnote{\href{https://qiskit.github.io/ecosystem/}{https://qiskit.github.io/ecosystem/}}. 
Qiskit Aer\footnote{\href{https://qiskit.github.io/qiskit-aer/index.html}{https://qiskit.github.io/qiskit-aer/index.html}} is Qiskit library with high-performance QC simulators and noise models. Some simulators included in Aer have support for leveraging Nvidia CPUs with Cuda version 11.2 or newer. Qiskit, Qiskit  Aer and Cuda relations in the development and execution environment are presented in Figure \ref{fig:executionstack}.

Nvidia CUDA\footnote{\href{https://developer.nvidia.com/cuda-zone}{https://developer.nvidia.com/cuda-zone}} is a computing platform developed for GPUs, for computationally demanding tasks suitable for parallel computing with up to thousands of threads. cuQuantum\footnote{\href{https://docs.nvidia.com/cuda/cuquantum/}{https://docs.nvidia.com/cuda/cuquantum/}} is an SDK based on CUDA, offering libraries for Quantum computing, with two libraries, cuStateVec for state vector computation and cuTensorNet for tensor network computation. cuStateVec is used by gate-based general quantum computer simulators, providing measurement, gate application, expectation value, sampler and state vector movement. CuStateVec library is available for Cuda versions 11 and 12. Nvidia cuQuantum is used by several different QDK's  GPU-powered quantum simulator backends. 

Developing across all target execution environments exposes the quantum software developer to a wide range of technologies that force them to balance their primary development activities with deep dives into operational aspects like configuring and maintaining their development environments or getting access to compatible hardware accelerators for running the relevant simulators. For example, Figure \ref{fig:executionstack} provides an overview of the software stack that application or algorithm developers using the Qiskit tools must be aware of. The situation is similar for other mainstream toolkits like PennyLane or Cirq\footnote{\href{https://quantumai.google/qsim/cirq\_interface}{https://quantumai.google/qsim/cirq\_interface}}. Experimental programming toolkits, like Eclipse Qrips \cite{qrisp}, leverage the existing Cirq or Qiskit assets to be able to execute circuits on GPU-accelerated simulators.
\subsection{Notebooks}

JupyterLab\footnote{\href{https://jupyter.org}{https://jupyter.org}} offers a versatile and user-friendly interactive computing platform suitable for data science, scientific computing, machine learning, and quantum computing. With its flexible architecture and extensive plugin ecosystem, it allows its users to develop customized workflows tailored to their specific needs, such as data exploration, prototyping algorithms or creating interactive presentations.

The key enabler of Jupyter is the notebook, an interactive and collaborative document formed by a collection of cells that can contain code, Markdown\footnote{\href{https://spec.commonmark.org/current/}{https://spec.commonmark.org/current/}} formatted text, equations or interactive widgets. A kernel is a computational engine that executes the code contained within the notebook. Jupyter supports multiple programming languages through different kernels, such as Python, R, Julia, and others. Users can select the desired kernel depending on their preferred programming language for a specific notebook. These combined capabilities allow scientists and algorithm developers to perform their work using a combination of code, explanatory text, and visualizations, making it easier to experiment, iterate, and document the development process.

\begin{figure}
    \centering
    \includegraphics[width=0.7\columnwidth]{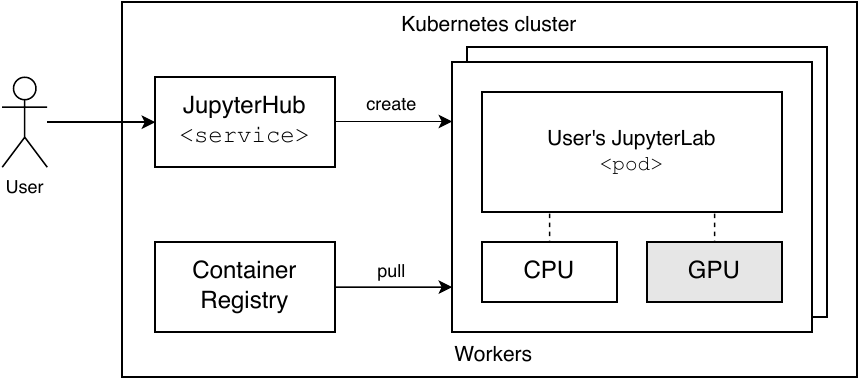}
    \caption{JupyterHub on Kubernetes architectural structure used in the mentioned solutions \textit{The Littlest JupyterHub} and \textit{Zero to JupyterHub for Kubernetes}.}
    \label{fig:jupyterhub}
\end{figure}

JupyterHub\footnote{\href{https://jupyter.org/hub}{https://jupyter.org/hub}} expands the functionality of JupyterLab to groups of users, giving them access to computational environments and resources without the burden of installation and maintenance tasks. The project provides two distributions: \textit{The Littlest JupyterHub} -- suitable for small group of users, typically less than 100, can be installed on a single virtual machine, and \textit{Zero to JupyterHub for Kubernetes}\footnote{\href{https://z2jh.jupyter.org/en/latest/index.html}{https://z2jh.jupyter.org/en/latest/index.html}} -- suitable for large number of user, makes extensive use of container technologies, cloud resources and infrastructure. Overview of arcitechtural structure described in Figure \ref{fig:jupyterhub}. The container that runs JupyterLab can be customised following the Jupyter Docker Stacks\footnote{\href{https://jupyter-docker-stacks.readthedocs.io/en/latest/index.html}{https://jupyter-docker-stacks.readthedocs.io/en/latest/index.html}} convention, allowing the user to run quantum algorithms in GPU accelerated simulators like Qiskit Aer or PennyLane Lightning. However, as the pod life cycle is linked to the user session, the GPU is locked by the user's pod regardless if the Python kernel executes code or not, a utilization pattern that is not optimal.

\subsection{Kubernetes for quantum}
Qubernetes \cite{qubernetes} (or Kubernetes for quantum) models the quantum computation tasks and the hardware capabilities required to execute them following established cloud-native principles, allowing seamless integration into the Kubernetes ecosystem. Following these conventions, a developer can submit quantum computation tasks packaged as jobs to Kubernetes clusters, which are executed on quantum capable nodes in simulators using classical computational resources (e.g. CPUs or GPUs), or on actual quantum hardware.

\subsection{Functionality}
The goal of developing the solution drives from the need for practical tools, for quantum software developers allowing to follow earlier presented SDLC's workflows. Following the model emphasizes the need for practical tools, enabling iterative workflow for quantum software development, with the possibility to transform execution from one platform to another when advancing in the process. Practically all modern software development methods advise towards iterative development and frequent code executions, to enable this to be done with quantum code, the execution needs to be as efficient as possible. 
Moving the execution from local to remote platform needs to provide a noticeable difference in execution efficiency to be beneficial for the developer.

\subsection{System architecture and components}

\begin{figure}
    \centering
    \includegraphics[width=\columnwidth]{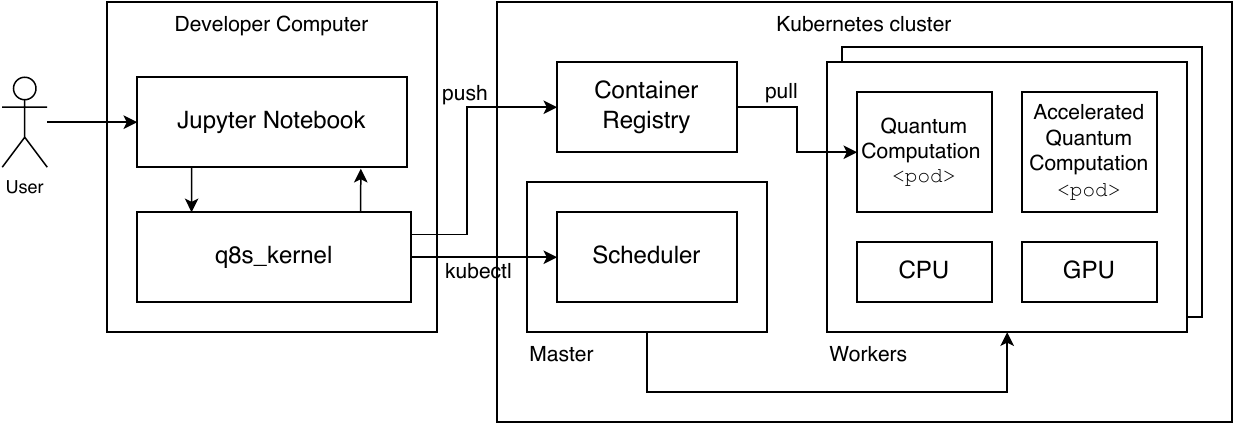}
    \caption{System architecture for the solution build to support developer in quantum software development. In the model introduced developer runs Jupyter lab locally outside of the Kubernetes cluster. }
    \label{fig:system}
\end{figure}
The solution enables a quantum software developer to run quantum routines or programs using GPU-accelerated simulators (e.g. Qiskit Aer or Pennylane Lightning) on a remote Kubernetes cluster with possibility to access more efficient computational resources, when comparing to local laptop execution.  The solution involves a custom Jupyter kernel (e.g., \texttt{q8s\_kernel}), and a compatible cluster that has at least one node that allows the execution of GPU-accelerated containers via the Nvidia Container Toolkit\footnote{\href{https://docs.nvidia.com/datacenter/cloud-native/container-toolkit/latest/index.html}{https://docs.nvidia.com/datacenter/cloud-native/container-toolkit/latest/index.html}}. The performance of the solution is related to the GPU's included in the cluster, and it is up scalable by applying more, or higher performing processing units to the cluster. To utilize the solution, the developer must install the kernel and specify the location of the configuration file of the cluster (e.g., \texttt{kubeconfig}\footnote{\href{https://kubernetes.io/docs/concepts/configuration/organize-cluster-access-kubeconfig/}{https://kubernetes.io/docs/concepts/configuration/organize-cluster-access-kubeconfig/}}) as an environment variable. The notebook is launched from the command line in Unix-based systems with simple command \texttt{KUBECONFIG=/path/to/kubeconfig jupyter lab}, where the \texttt{"/path/to/kubeconfig"} is replaced with the actual path to the cluster's configuration file, making the startup of the kernel and accessing the GPU in the cluster as simple as possible for the developer. Through the user interface of the Jupyter Notebook/Lab, the user can switch between the local development kernel (e.g. \texttt{IPython}) and the remote Kubernetes cluster. The system architecture and components are detailed in Fig.~\ref{fig:system}. As the Jupyter notebook leveraging the solution is run self-hosted, the developer is able to access any compatible notebook they have stored locally, and use their personalized settings or extensions in the Jupyter, as they would when working in a local development environment. The Kubernetes cluster uses a container base image including the necessary quantum libraries, and other dependencies related to quantum execution using GPUs.

\subsection{Task execution model}

\begin{figure*}[!t]
    \centering
    \includegraphics[width=1\textwidth]{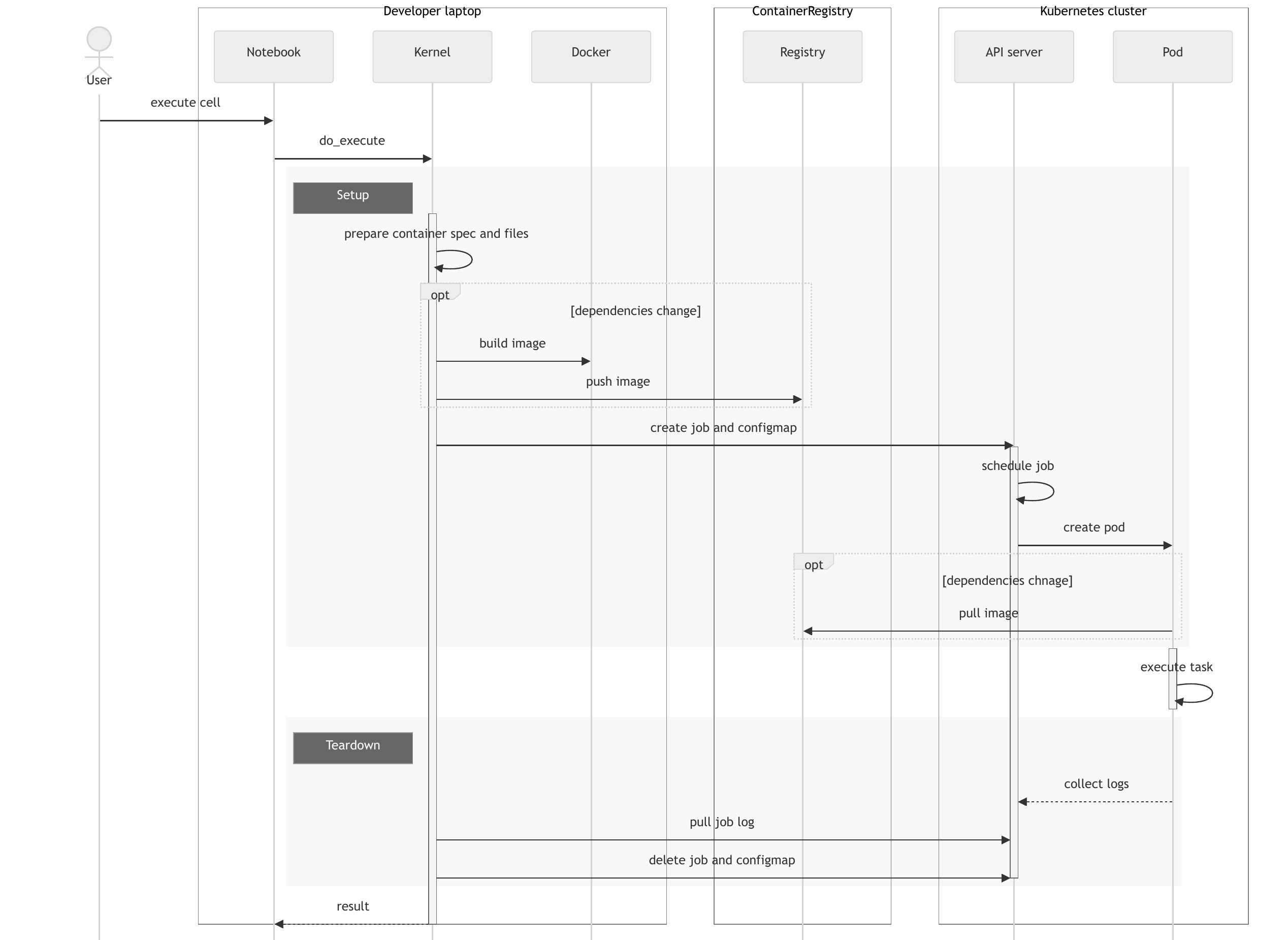}
    \caption{Execution flow of a notebook cell on Kubernetes cluster using the q8s\_kernel.}
    \label{fig:seq}
\end{figure*}

The execution flow is triggered by the user pressing the \textit{run} button in the notebook. When the kernel receives the \texttt{do\_execute} command, it detects the dependencies in the cell code and prepares the container specification (e.g., \texttt{Dockerfile} and \texttt{requirements.txt}), using as base image a pre-build image that includes all dependencies for the CUDA version supported in the cluster. The kernel builds the image and pushes it to the container registry. Then it creates a Kubernetes Job specification that corresponds to the execution task (see Listing \ref{listing:job}), and a ConfigMap that contains the actual code that will be mounted as a volume in the Pod. Once the cluster API server receives the request, it schedules the job when the requested GPU resources are available. The Pod pulls the image from the Registry and executes the tasks. The kernel polls the API server for the Job's status waiting for completion, then collects the logs and cleans up by deleting the Job and the ConfigMap. Depending on the container's exit code (e.g. success for 0, or failure otherwise), the kernel returns the result to the notebook on the \texttt{stdout} or \texttt{stderr} respectively. The kernel rebuilds the image and the pod pulls the image only when dependencies change. The task execution sequence is depicted in Fig. \ref{fig:seq}.

\begin{listing}[t]
\begin{minted}[fontsize=\scriptsize,xleftmargin=1em,linenos=true,highlightlines={12-13,16,19,23}]{yaml}
apiVersion: batch/v1
kind: Job
metadata:
  name: "quantum-job"
spec:
  template:
    metadata:
      name: "quantum-pod"
    spec:
      containers:
        - name: "quantum-task"
          image: registry.com/user/job-dependencies:v1
          command: ["python", "/app/main.py"]
          resources:
            limits:
              nvidia.com/gpu: '1' # requires GPU usage
          volumeMounts:
          - name: source-code-volume
            mountPath: /app
      volumes:
        - name: source-code-volume
          configMap:
            name: task-files #{"main.py": "code"}
      restartPolicy: Never
\end{minted}
\caption{Quantum job specification}
\label{listing:job}
\end{listing}

\section{Evaluation}
\label{sec:evaluation}

\subsection{Benchmark scenarios}

We have evaluated the solution in the following scenarios that we consider representative of how the solution will be used. The baseline consists of the user running the development environment (e.g. the Jupyter Notebook/Lab) and executing the quantum routine experiments on his own laptop. The following test scenarios employ CUDA capable GPUs accessed remotely in Kubernetes clusters:

\textbf{Cluster with mobile workstation} - Users with better hardware share their computational resources (e.g. a mobile workstation) with the rest of the team in a Kubernetes cluster. Users run the development environment similar to the baseline scenario, but the quantum routines are executed on the mobile workstation. The cluster is not used by other users while the benchmark routines are executed.

\textbf{Cluster with cloud GPUs} - The user runs the development environment on his own laptop and executes the quantum routine experiments on a Kubernetes cluster operated by a commercial entity. In our case, we have selected Puzl\footnote{\href{https://puzl.cloud/}{https://puzl.cloud/}}, a provider that offers access to Nvidia A100 40GB GPUs. The cost of using the GPU resources is approximately 1.6 EUR/h, in line with other cloud infrastructure providers. The charging model is based on effective utilization of the GPU resource, e.g., the effective time the Job runs to completion. The cluster is shared with the other Puzl users that execute their own workloads while our benchmark routines are executed.

The detailed hardware configurations of the devices used in the evaluation scenarios are described in Table~\ref{tab:hardware}.

\begin{table*}
    \caption{Hardware and software configurations used for the benchmark environment.}
    \label{tab:hardware}
    \centering
    \begin{tabular}{p{2.9cm}p{2.4cm}p{3cm}p{3cm}}
    \hline
    \bfseries Scenario & \bfseries Baseline & \bfseries Mobile Workstation & \bfseries Cloud GPU \\
    \hline
    Hardware category & Business laptop & Mobile workstation & Cloud server \\
    Model/Provider & Dell Latitude 7440 & Dell Precision 7680 & puzl.cloud \\
    CPU & Intel i5-1345U 16GB & Intel i9-13950HX 64GB & 2 vCPUs up to 64GB \\
    GPU (CUDA compatible) & - & Nvidia GeForce RTX 4090 Laptop 16GB & Nvidia A100 40GB \\
    GPU driver& - & CUDA 12.2 & CUDA 12.2 \\
    Host OS& Windows 10& Ubuntu 22.04 & -\\ 
    Guest OS& - & Ubuntu 22.04 & Ubuntu 22.04 \\ 
    Python version &Python 3.10 &Python 3.10 &Python 3.10 \\
    Qiskit version & 1.0.0& 1.0.0& 1.0.0\\
    Qiskit AER & 0.13.3 & 0.13.3 & 0.13.3\\
    
    \hline
    \end{tabular}
\end{table*}

\subsection{Benchmark tooling}

\begin{listing}[t]
\begin{minted}[fontsize=\scriptsize,linenos=true,xleftmargin=1em,highlightlines={},highlightcolor
=lightgray]{python}
import timeit
from testbook import testbook

QUBIT_START = 3
QUBIT_END = 29
ITERATIONS = 10

def benckmark(notebook, kernel_name, device, target):
  for qubits in range(QUBIT_START, QUBIT_END + 1):
    for iteration in range(1, ITERATIONS + 1):
      @testbook(
        notebook,
        execute=True,
        kernel_name=kernel_name
      )
      def test(tb):
        start = timeit.default_timer()
        func = tb.get("test_function")

        simulator = func(
          qubits, 
          device=device, 
          target=target
        )
        end = timeit.default_timer()
        overhead = end - start - simulator
        # log test results

      test()
\end{minted}
\caption{Template for the benchmark script.}
\label{listing:benchmark}
\end{listing}

We developed a benchmarking tool based on \texttt{testbook}\footnote{\href{https://pypi.org/project/testbook/}{https://pypi.org/project/testbook/}}, a unit testing framework for testing code in Jupyter Notebooks. The tool implements the following workflow (see Listing \ref{listing:benchmark}): loads the notebook containing the test function, configures the notebook with a specific kernel -- \texttt{Python} for local execution and \texttt{q8s\_kernel} for remote execution on Kubernetes cluster, and invokes the test function (see Listing \ref{listing:program}), with number of qubits and target device -- \texttt{CPU} for local test or \texttt{GPU} for Kubernetes. Each test run measures the \textit{simulator} time -- the amount of time spent executing the test function in the simulator, and the \textit{overhead} time -- the amount of time required for interacting with the kernel (e.g. local tests), or the time required to setup and teardown the job that executes the computation task in the Kubernetes cluster. The test procedure is repeated 10~times and the resulting values are averaged.

\begin{listing}
\begin{minted}[fontsize=\scriptsize,linenos=true,xleftmargin=1em,highlightlines={},highlightcolor
=lightgray]{python}
import timeit
from qiskit import transpile
from qiskit.circuit.library import QFT as Test
from qiskit.circuit.library import QuantumVolume as Test
from qiskit.circuit.library import QAOAAnsatz as Test
from qiskit_aer import AerSimulator
from qiskit.transpiler import CouplingMap
from qiskit_aer.noise import NoiseModel
from qiskit_ibm_runtime.fake_provider import FakeAuckland

def test_function(n, method="statevector", device='GPU')
    cm = CouplingMap().from_full(n)
    model = FakeAuckland()
    noise_model = NoiseModel.from_backend(model)
    backend = AerSimulator(
        noise_model=noise_model,
        method=method, 
        device=device, 
        coupling_map=cm
    )

    # One of QFT, QuantumVolume, QAOAAnsatz
    circuit = Test(num_qubits=n)
    circuit.save_state()
    circuit = transpile(
        circuit, 
        backend=backend, 
        coupling_map=cm
    )

    start = timeit.default_timer()
    backend.run(circuit).result()
    end = timeit.default_timer()
    
    # simulator value in benchmark script
    return end - start
\end{minted}
\caption{Template for the quantum benchmark routines}
\label{listing:program}
\end{listing}

\subsection{Quantum benchmark routines}

\begin{figure*}[!t]
    \centering
    \subfloat[]{\label{speedup-qft}
    \includegraphics[width=0.5\textwidth]{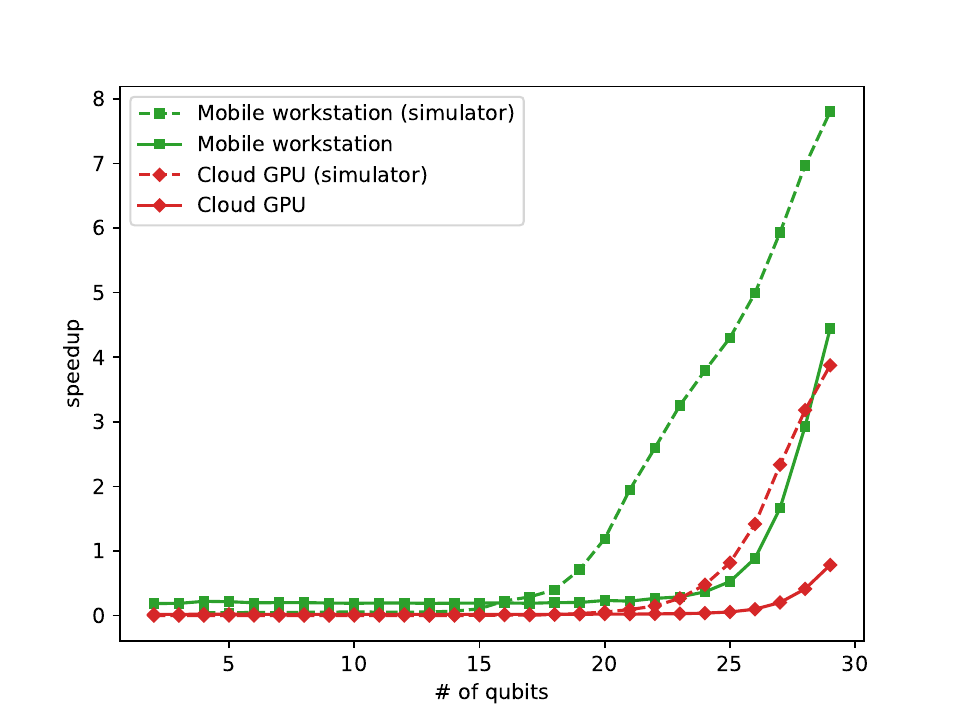}
    }
    \subfloat[]{\label{speedup-qv}
    \includegraphics[width=0.5\textwidth]{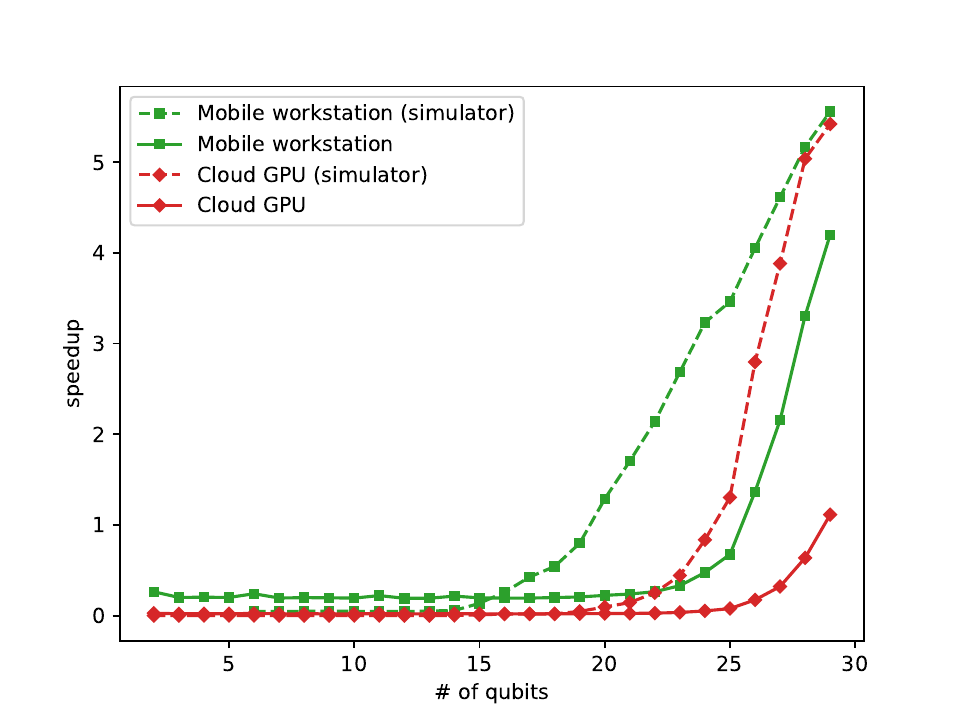}
    }
    \hfill
    \subfloat[]{\label{speedup-qaoa}
    \includegraphics[width=0.5\textwidth]{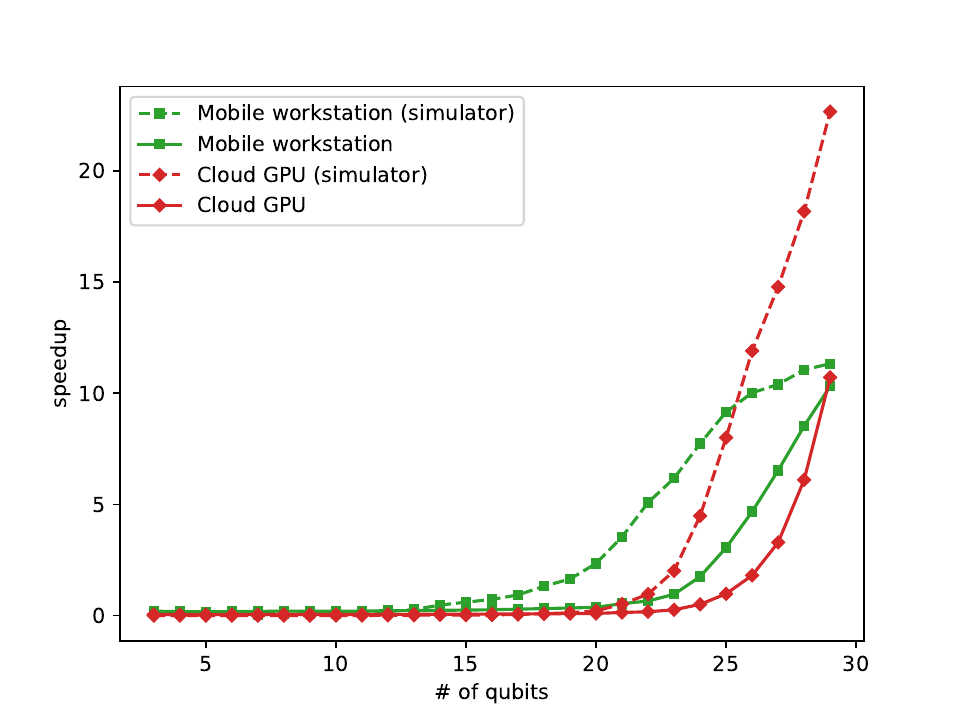}
    }
    \caption{Quantum test routines execution speedups: QFT (a), QV (b) and QAOA (c). The baseline for the speedup is local execution, with laptop and CPU, The dotted line represents the maximum speedup that can be achieved with the raw processing power of the GPU, whereas solid lines correspond to the actual speedup that considers also the network overhead needed to set up and tear down the remote routine execution.}
    \label{fig:speedups}
\end{figure*}

\begin{figure*}[!t]
    \centering
    \subfloat[]{\label{alg1-a}\includegraphics[width=0.5\textwidth]{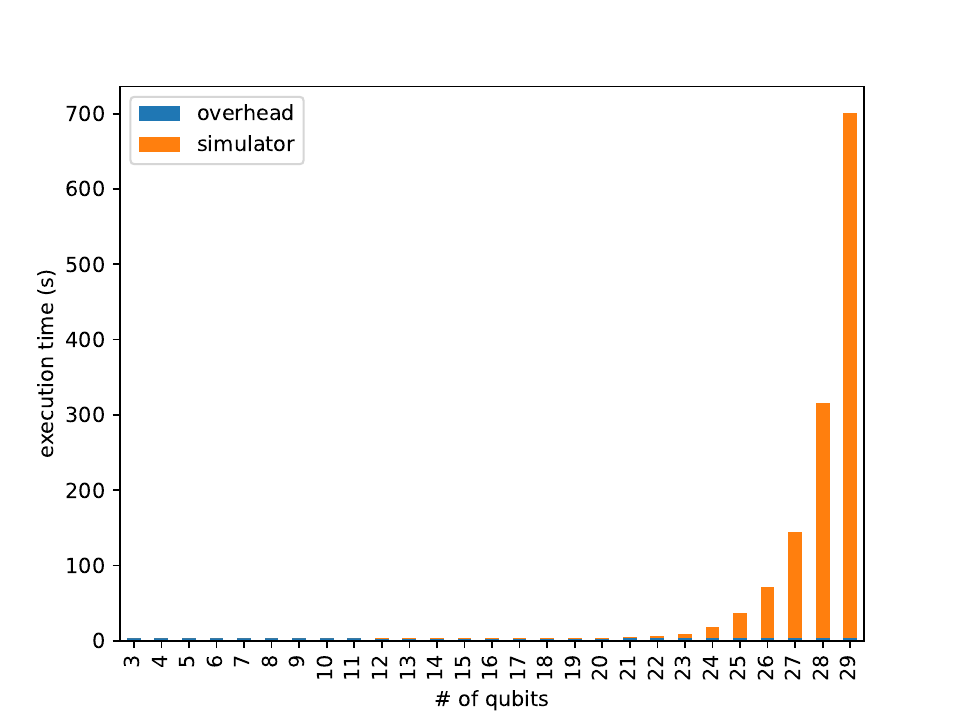}}
    \subfloat[]{\label{alg1-b}\includegraphics[width=0.5\textwidth]{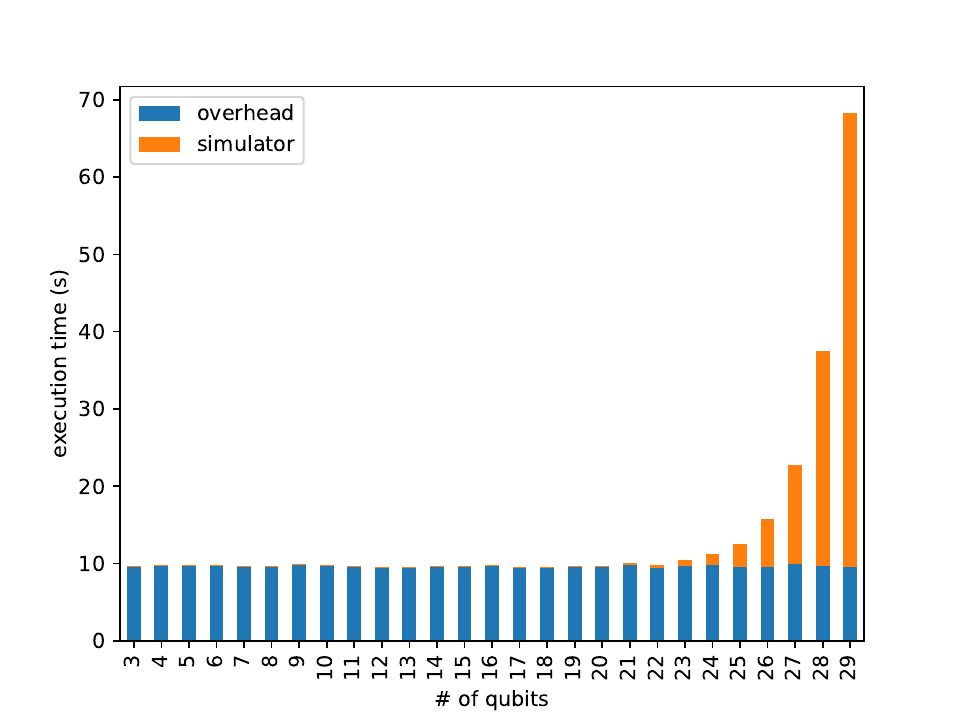}}
    \hfill
    \subfloat[]{\label{alg1-c}\includegraphics[width=0.5\textwidth]{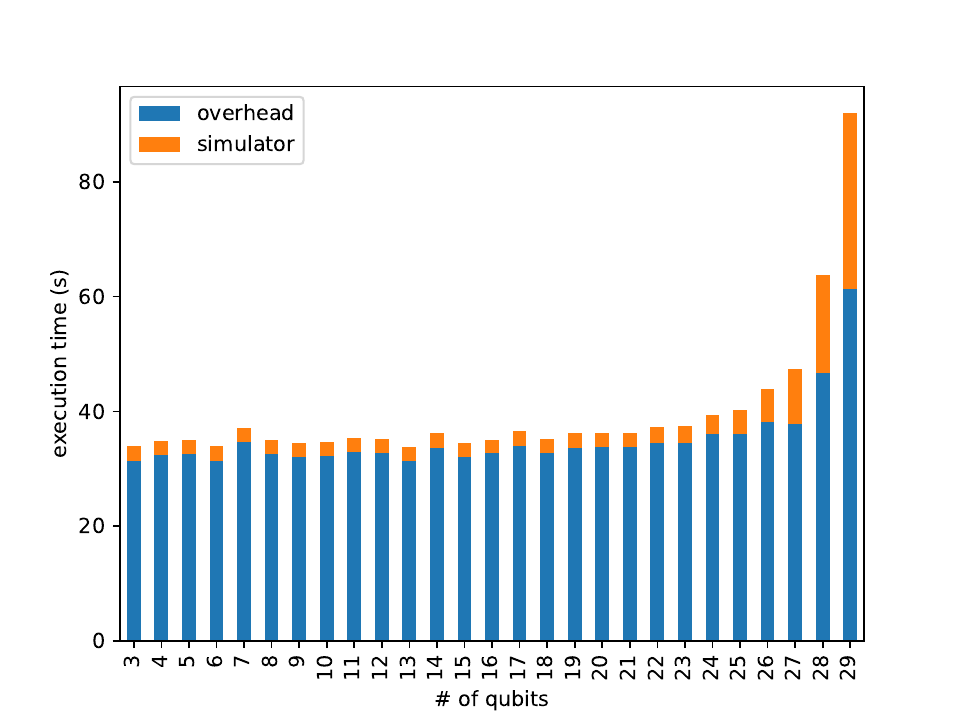}}
    \caption{Execution time for the QAOA test routine in baseline scenario (a), mobile workstation scenario (b), and cloud GPU scenario (c).}
    \label{fig:QAOA}
\end{figure*}

\begin{figure*}[!t]
    \centering
    \subfloat[]{\label{alg1-a}\includegraphics[width=0.5\textwidth]{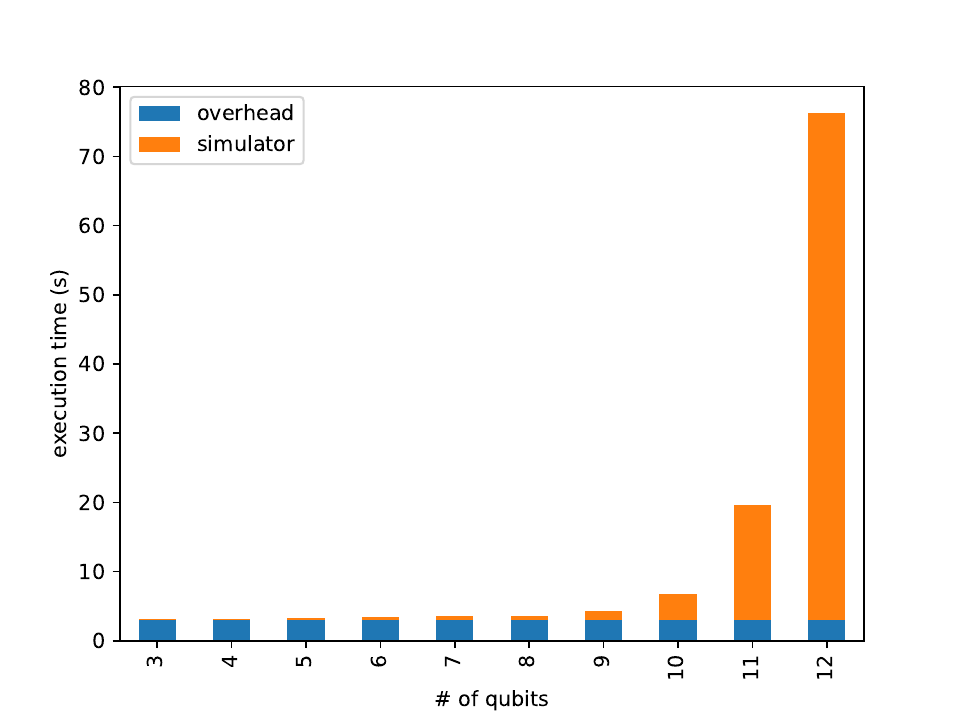}}
    \hfill
    \subfloat[]{\label{alg1-b}\includegraphics[width=0.5\textwidth]{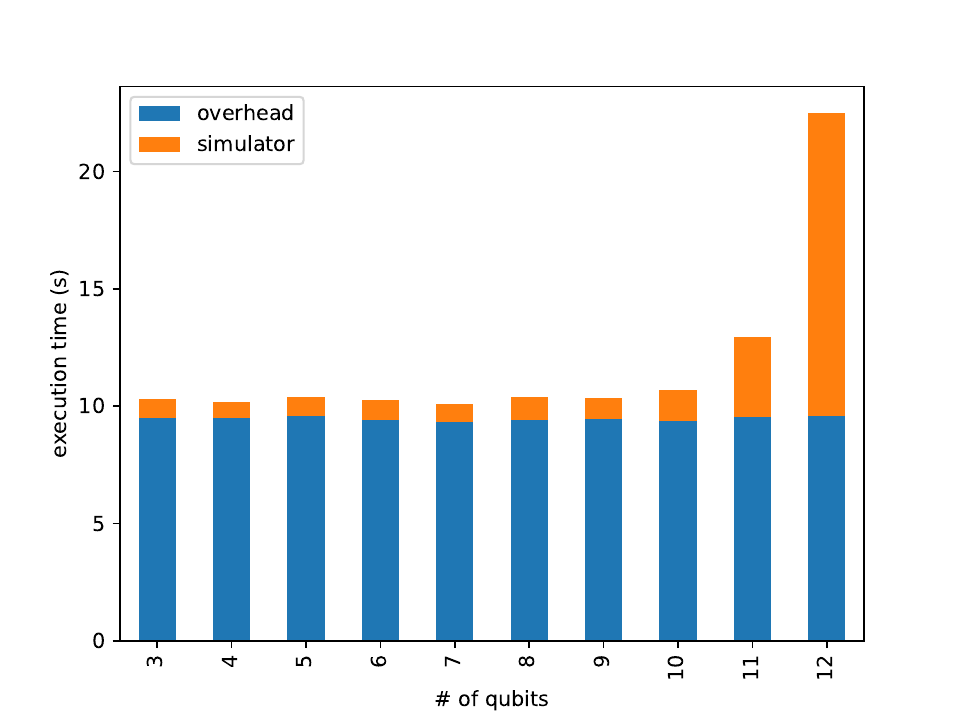}}
    \hfill
    \subfloat[]{\label{alg1-c}\includegraphics[width=0.5\textwidth]{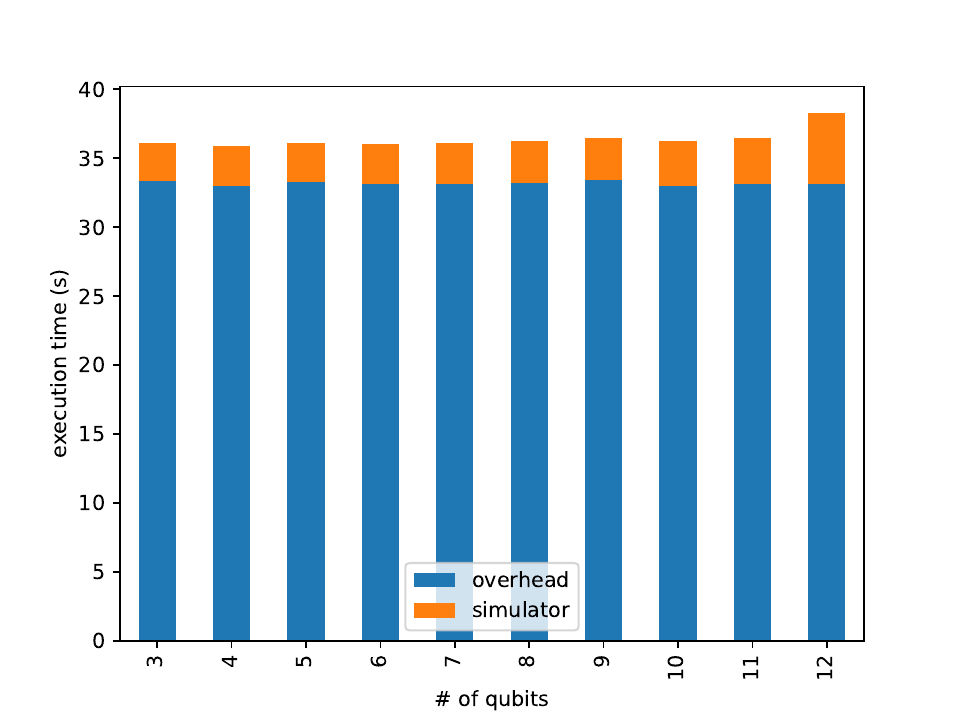}}
    \subfloat[]{\label{alg1-c}\includegraphics[width=0.5\textwidth]{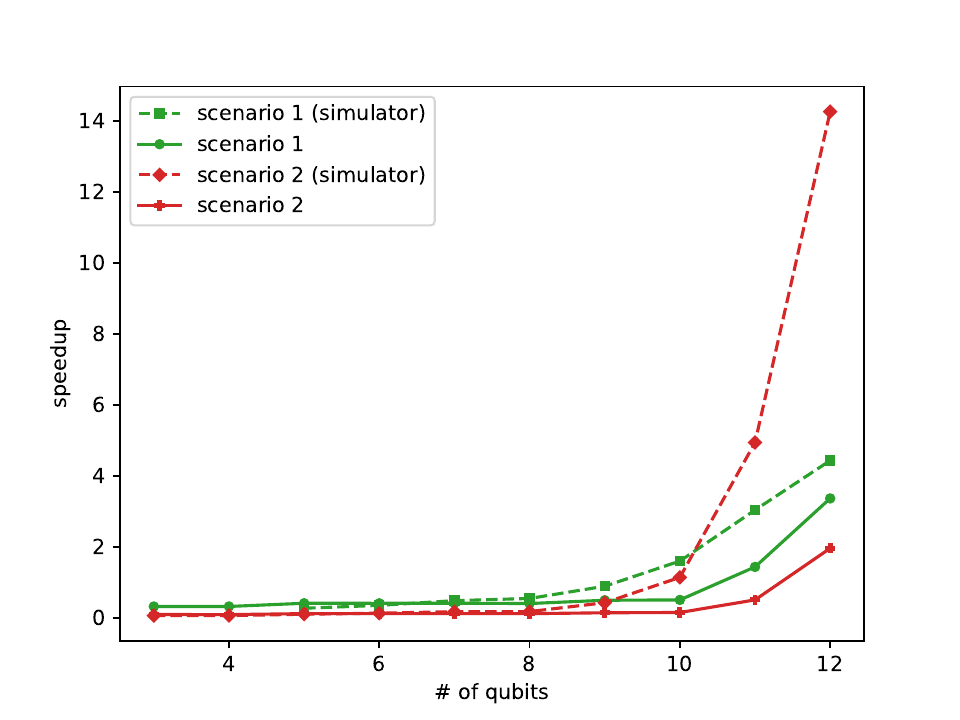}}
    \caption{QAOA with noise test routine in baseline scenario (a), mobile workstation scenario (b), cloud GPU scenario (c), and speedup comparison (d).}
    \label{fig:QAOA-noise}
\end{figure*}

For each benchmark scenario, we tested three quantum routines: the Quantum Fourier Transform (QFT) circuit, the Quantum Volume (QV) metric~\cite{QV} and the Quantum Approximate Optimization Algorithm (QAOA) for the Max-Cut problem~\cite{QAOA}. Similar selection of quantum routines have been used in benchmarking focusing on software, \cite{cuquantumbenchmark, giortamisorchestratingquantumcloudenvironments}. The selection of algorithms might differ from a selection of benchmarking used for QPU hardware and/or its components benchmarking, some of which focus on phenomena like error rates, that are not present when running simulators.

\textbf{QFT} is a key element of many fault-tolerant quantum algorithms like Shor's algorithm~\cite{shor} and the Harrow–Hassidim–Lloyd (HHL) algorithm~\cite{hhl}, which show a theoretical exponential advantage over their classical counterparts. Fault-tolerant quantum computing is not likely to be achievable in the short term, but as better quantum error correction (QEC) techniques and better hardware become available, a reliable and quick way to simulate circuits like the QFT is important for future benchmarks. Therefore, it is natural to choose this routine as the fault-tolerant benchmark scenario.

\textbf{QV} is a single-number metric that quantifies the largest random circuit of equal depth and width that a given quantum computer can implement successfully, up to an effective error rate. The QV is calculated from a circuit of $d$ layers of two-qubit unitary gates sampled from the Haar measure on SU(4) applied to random partitions of pairs of qubits. The QV can also be understood as the complexity of simulating this random circuit on classical computers, so it functions as a good benchmark example for our tests.

\textbf{QAOA}. As current quantum hardware has a limited number of qubits and suffers from noisy gates and poor coherence times, variational quantum algorithms \cite{Cerezo2021} emerge as a promising alternative to achieve quantum advantage, combining the power of QPU and classical optimization algorithms. Therefore, studying the performance of classical simulators for these kinds of algorithms is important to further understand the limitations of hybrid classical-quantum approaches. One of the most popular variational algorithms in the NISQ era is the QAOA, a hybrid quantum-classical algorithm for solving optimization problems. In QAOA, a parameterized quantum state is prepared, that maximises the cost function of the corresponding optimization problem, using $p$ layers of parameterized unitaries. 
\\

The main objective of these benchmark tests is to compare the execution time between the different scenarios, as the number of qubits and gates within circuits grows larger. We performed several circuit simulations for each quantum routine using Qiskit for circuits with varying numbers of qubits up to 29 qubits. We simulated the exact QFT circuit, and simulated QV circuits with $d=20$ layers of random gates. For the QAOA algorithm, we consider the Hamiltonian of the Max-Cut problem of a 2-regular graph on $n$ nodes, where $n$ corresponds to the number of qubits. We simulated the QAOA circuit using $p=5$ layers of cost and mixer Hamiltonians and random initial parameters. As we are only interested in the execution time of the simulator, we ignore the classical optimization loop and focus only on simulating the quantum circuit. Using a fully connected coupling map, the number of gates of each routine scales as $\mathcal{O}(n(n/2 + 1))$ for QFT, $\mathcal{O}(dn/2)$ for QV and $\mathcal{O}(pn(n+ 1) + n)$ for QAOA.

Until now, we have only considered ideal circuits, but practical benchmarks also require including realistic noise models and coupling maps to obtain results closer to experimental results. For that reason, we also performed the same QAOA benchmark including a noise model, coupling map and basis gates set taken from \texttt{FakeAuckland}, a 27 qubit backend available in qiskit-ibm-runtime\footnote{\href{https://pypi.org/project/qiskit-ibm-runtime/}{https://pypi.org/project/qiskit-ibm-runtime/}}.

\subsection{Execution speedup}

We successfully executed selected QFT, QV, and QAOA test routines in circuits containing up to 29 qubits in the baseline and two test scenarios. Despite previous reports suggesting that circuits with 31 qubits require 17GB of GPU RAM~\cite{faj2023benchmark}, our attempts to run 30 qubit circuits failed in both test scenarios. Although the jobs were scheduled and the pods started on the proper cluster node, they were terminated due to running out of memory. Qiskit Aer raised an error for circuits with 31 qubits, indicating before starting the simulation that the minimum GPU RAM requirements were not met.

The results, illustrated in Fig. \ref{fig:speedups}, indicate that speedups begin to emerge for all quantum routines when circuits exceed 24 qubits. Both QFT and QV routines exhibit similar speedup patterns, despite the cloud GPU scenario having a more capable GPU than the one in the mobile workstation. However, the overhead of securing the necessary computational resources when competing with other users in the cluster outweighs the speed-up gains of the more powerful GPU. In contrast, the QAOA routine demonstrates significant speedups in both test scenarios, but a substantial advantage in the cloud GPU scenario when considering only the simulator execution time. This speedup can be attributed to the differences in the number of gates between QAOA and the other methods. For 29 qubits, the number of gates for QFT, QV and QAOA are 450, 280 and 4379, respectively.

The execution time details for the QAOA test routine are depicted in Fig. \ref{fig:QAOA}. We can see that for circuits larger than 24 qubits in the baseline scenario, the execution time exceeds the expected limits of a fast iteration read–eval–print loop (REPL~\cite{vanBinsbergen2020repl}) environment like Jupyter Notebook, shifting to batch execution mode. In both mobile workstation and cloud GPU testing environments, the execution time follows the same exponential growth pattern, but with smaller values, resulting in a 10x speedup for 29-qubit circuits. The overhead in the mobile workstation scenario remains relatively constant due to available computational resources. In contrast, the cloud GPU scenario shows increasing overhead starting from 27 qubits, due to increased RAM requirements that the cluster scheduler needs to secure before allowing the execution. As a result, we can expect that although the overhead will increase in clusters with more capable GPUs, it will be at a slower rate than simulation time growth, thus not negatively impacting speedup gains.

The QAOA routine was executed on the backend with \texttt{FakeAuckland} noise model up to circuits with a maximum of 12 qubits, see Fig. \ref{fig:QAOA-noise}. This was the highest circuit width that could be run on both our mobile workstation and cloud GPU testing environments, as larger circuits exceeded the available GPU's RAM capacity. Despite the relatively small number of qubits in our experiments, we observed speedup benefits starting at around 10 qubits. Although our findings are not definitive, we predict that GPUs with more memory would be able to execute circuits on noisy backends with significant speedup.

The quantum benchmark routines used in the evaluation were selected considering the increased computational capabilities required to execute circuits with a larger number of qubits and an increasing number of gates, which ultimately convert into longer execution times. The results in both test scenarios demonstrate that significant speedups for circuits larger than 25 qubits, allow quantum software developers to perform experiments by iterating faster, which ultimately improves their productivity. Thus, the objective O\ref{obj:speedup} is achieved.

\subsection{Improving iterative development with tooling}

\begin{figure}
    \centering
    \includegraphics[width=\columnwidth]{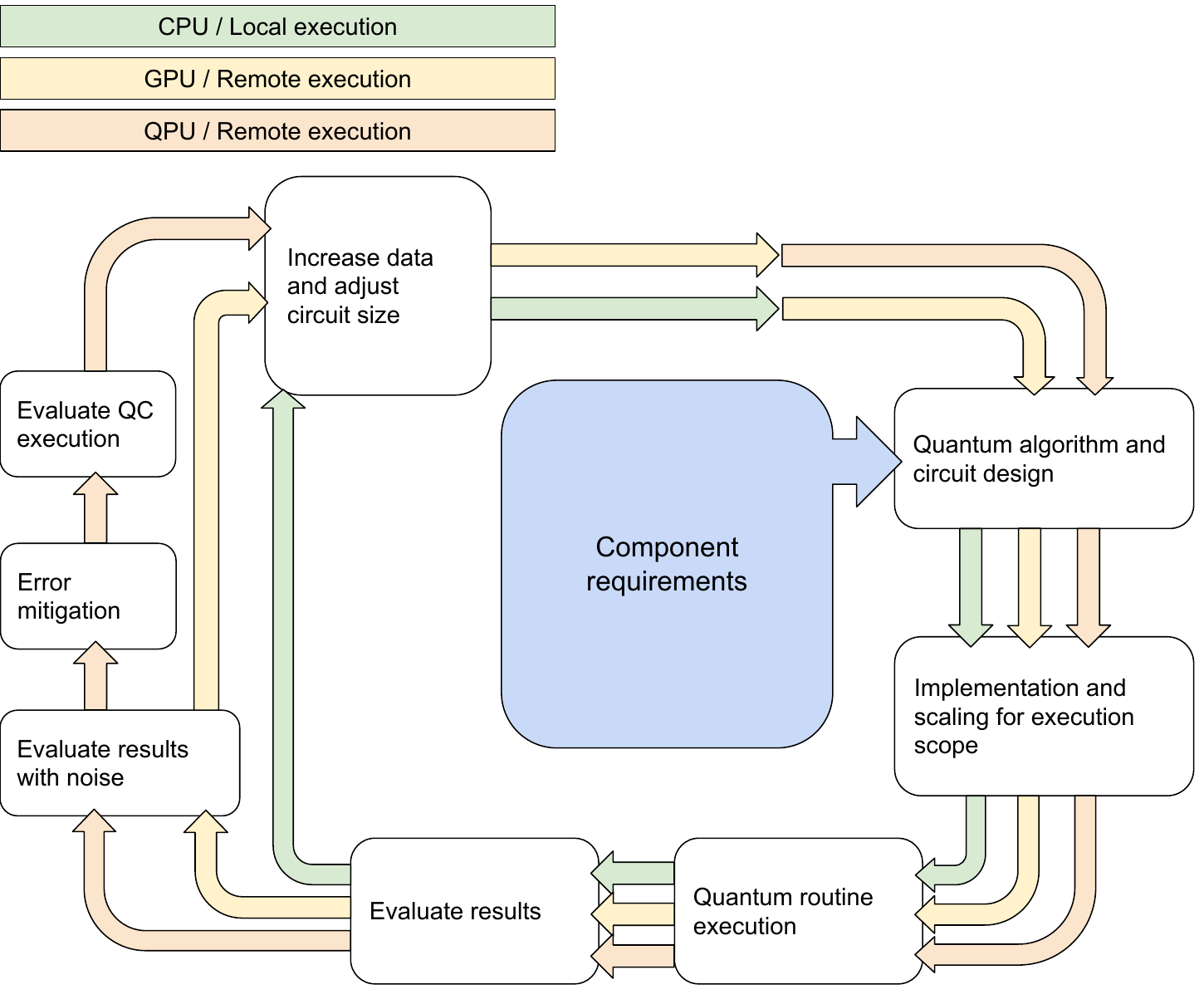}
    \caption{Workflow for quantum software and quantum circuit development with iterative model. The suggested flow stars from requirements and follows an incremental and iterative flow, moving from local environment to more efficient and remote platforms on each iteration.}
    \label{fig:workflow}
\end{figure}

\

Our take on the workflow for quantum software development, presented in Fig. \ref{fig:workflow}, follows the earlier presented guidelines by SDLC and ICSM and focuses on clarifying changes in the environment during the process. The process starts from the requirements for the quantum components, then followed by the algorithm and circuit design on selected quantum SDK. Moving forward to implementation of the circuit on scale for the current execution target, then to execution on the selected hardware, locally in the earlier stages and incrementally moving to remote GPU and later to QPU. The events that necessitate a change in environment generally fall into two categories: 1) inefficient execution time with a simulator, and 2) memory limitations of the execution platform. Among these, we considered it to be almost as important to have a possibility to iterate back temporarily from QPU to GPU,  e.g. when fine tuning algorithms and data, as repeating execution will noticeably increase the running cost. Following the execution phase, in all cycles the results should be evaluated, and in the later iterations introduced and evaluated with the noise on the results. Depending on the quality of the results, the input data and the circuit size should be adjusted when moving to the next iterations.

\begin{figure}
    \centering
    \includegraphics[width=\columnwidth]{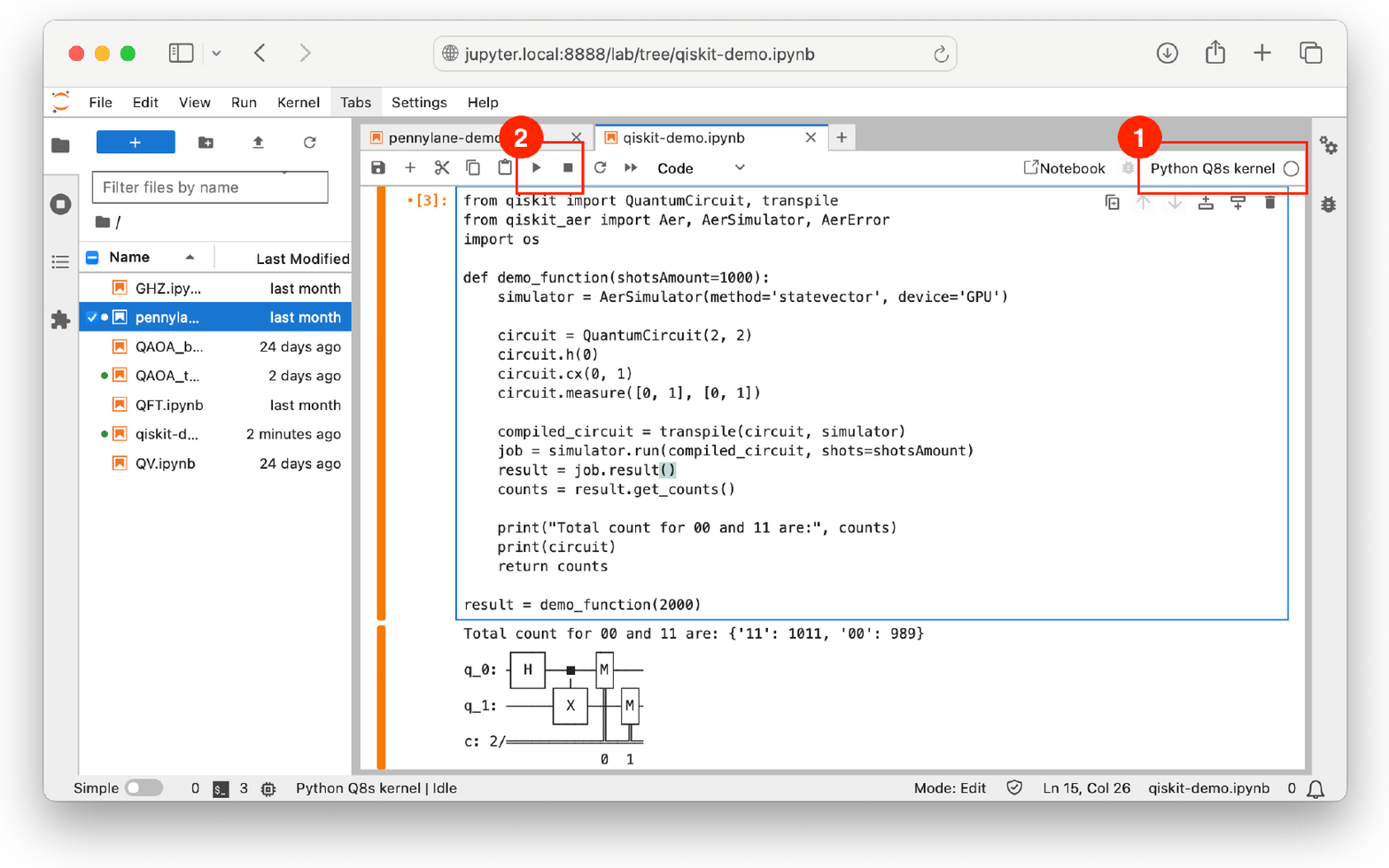}
    \caption{Jupyter lab environment configured for executing quantum computations on a remote cluster via the "Python Q8s kernel" (q8s\_kernel). Target execution environment can be switched using the kernel selection capability (1), while the execution can be started on the selected environment using the build in start/stop toolbar (2)}
    \label{fig:screenshot}
\end{figure}

The tooling to support the model is perceived by the user as a standard Jupyter kernel, see Fig. \ref{fig:screenshot}. To function properly, the implementation relies on Docker and Kubernetes, widely used tools supported on a multitude of operating systems. The selection of the cluster where the quantum task execution is performed is achieved by providing the \texttt{kubeconfig} configuration file as an environment variable. The solution does not require a deep understanding of Kubernetes cluster management beyond the configuration file. As such the user is not exposed to the complexities of enabling access to the GPUs or configuring the computational layer of the CUDA or cuQauntum. Leveraging the Jupyter kernel abstraction, the developers can switch the target execution to reflect the development stage they are focused on.

From the development workflow the perspective, the kernel works as enabler for the user in up-scaling the execution environment between the iterations in the development process. With the current implementation offering a easy access to local CPU and remote GPU from the same Notebook interface, all from developer's local environment. Enabling to perform fast iterations as the more computationally intensive quantum tasks are executed in less time using the remote computational resources of the Qubernetes cluster.
The objective O\ref{obj:iteration} and O\ref{obj:target} is achieved.
\subsection{Limitations}
The current implementation state of the tooling is limited by the lack of support for QPU and multi-GPU, which limits the applicability to a subset of the SDLC workflow, or executing routines that are bound by limits of one GPU instance.
Both represent future research directions and we address the further in section \ref{sec:conclusion}.
Nevertheless, the extendability ability of the tooling, consisting of the Jupyter kernel and the Qubernetes cluster has been considered from the beginning of the design process. With the proposed solution, the circuit execution is initiated by the kernel, which packages the quantum routine into a job that is annotated to match the target selected by the user. The Kubernetes scheduler component inside the cluster initiates execution on the available computational resource that matches the user request. With suitable QPU hardware integration inside the cluster visible as a node advertising computational capacity of \texttt{vendor.example.com/qpu: '1'}, switching the execution target is equivalent with changing \texttt{resources.limits} property in the Kubernetes job description. The precise changes between a GPU target and QPU target is presented in Table~\ref{tab:target}.

\begin{table}[]
    \centering
    \begin{tabular}{|l|l|} \hline 
 Target platform & Kubernetes job description\\ \hline 
    Remote GPU &         
    \begin{tabularlstlisting}
containers:
    - name: "quantum-task"
      image: registry.com/user/job-dependencies:v1
      command: ["python", "/app/main.py"]
      resources:
        limits:
          nvidia.com/gpu: '1' # requires GPU usage
    \end{tabularlstlisting}\\ \hline 
    Remote QPU &         
    \begin{tabularlstlisting}
containers:
    - name: "quantum-task"
      image: registry.com/user/job-dependencies:v1
      command: ["python", "/app/main.py"]
      resources:
        limits:
          vendor.example.com/qpu: '1' 
    \end{tabularlstlisting}\\ \hline
    \end{tabular}
    \caption{Kubernetes job description examples with GPU target and QPU target.}
    \label{tab:target}
\end{table}

\subsection{Threats to validity}
\label{sec:threats}

The threats to the validity of our study are discussed following to the categorization provided by Wholin et al.~\cite{Wohlin2012}, dividing the evaluation of validity to four areas, internal validity, external validity, construct validity and conclusion validity.

A threat to internal validity may arise from the selection of the quantum routines used for benchmarking, might not be representative of all development situations. To mitigate this threat, we have utilized a set of algorithms and routines found in other benchmarking experiments performed by academia~\cite{faj2023benchmark,jamadagni2024benchmarking} and industry~\cite{nvidia2022benchmarking}. Another threat to internal validity could arise from developing the benchmark experiments using only the Qiskit toolkit and executing the routines in the Qiskit Aer simulator. The mitigation, in this case, is that the speedups are determined to a large extent by CUDA and cuQauntum toolkits, which are used by other popular simulators, e.g., \texttt{lightning.gpu}\footnote{\href{https://docs.pennylane.ai/projects/lightning/en/stable/index.html}{https://docs.pennylane.ai/projects/lightning/en/stable/index.html}} for PennyLane or \texttt{qsimcirq}\footnote{\href{https://docs.nvidia.com/cuda/cuquantum/latest/appliance/cirq.html}{https://docs.nvidia.com/cuda/cuquantum/latest/appliance/cirq.html}} for Cirq.

A threat to our study's external validity arises from the performance objectives employed by operators of different Kubernetes clusters, which may reflect in lack of significant speedups on executing quantum routines due to the availability of the required computational resources (e.g. memory or GPUs). To mitigate this threat, we have used two Kubernetes clusters, one operated by us, and one that is a live system operated by a commercial entity focused on providing Nvidia A100 computing resources. Together, they allowed us to observe that even when relying on the Kubernetes built-in scheduling infrastructure, we are still able to observe significant speedups when executing quantum computations with circuits having more than 25 qubits. As we enroll actual quantum algorithm developers in our university's test environment, we will gain additional insights into how the cluster resources can be better utilized in order to reduce the overhead of executing quantum tasks in a cluster.

A threat to construct validity arises from the selection of Jupyter Notebook as the programming modality, which leaves out the developers that use text editors or integrated development (IDE) to write the quantum routines directly as Python files. The decision to focus on notebooks is that developers who prefer this programming metaphor are less likely to be knowledgeable about setting up and maintaining complex development environments. An approach to mitigate this threat is to split the functionality of the q8s\_kernel into the Jupyter and the Kubernetes specific components and extract the latter into a command line interface (CLI) tool that can be used independently.

As a threat to conclusion validity, we recognize that for some quantum software developers there are other options, such as having direct access to local development environment with high powered infrastructure, or high availability on resources to run executions on QPUs. Furthermore, with a certain level of expertise in software engineering, the developers have a possibility to set up their own infrastructure, built with different components, to reach similar results.

\section{Conclusions and future work}
\label{sec:conclusion}

In the related work focusing on quantum software development practices, SDLC  for quantum\cite{sdlc} and ICSM for quantum\cite{perezguidelines}, the authors propose models with a specific focus on the development process. In this work we propose a software tooling centered around a Jupyter kernel and Qubernetes that enables practitioners to implement and follow the suggested development models in practice. As the result, we present workflow model with apractice for scaling up the execution platform from local environment to efficient remote platform, and finally to quantum hardware, and the tooling required to support the proposed practice. The results emphasize that the quantum circuit simulation is an important part of quantum programming in the near future, although ultimately the finalized program and algorithms should be executed on an actual QPU. We have built the solution on versatile platforms, the Kubernetes cluster used in the solution, to schedule and deliver the workloads is suitable to handle quantum workloads with simulators as well as quantum hardware, and the Jupyter kernel packaging the code for execution does not make any difference on which platform the code is finally executed. 

The natural next topic to research following this contribution will be the effort of integrating a suitable QPU into the cluster, and into the kernel for the Jupyter notebook. This would further benefit the developer by enabling the iterative approach to cover all stages of the development process, using the familiar notebook environment. QPU integration in the presented system would open up the possibility also for further research and usage of the solution on quantum-classical hybrid algorithms execution, by offering access to both efficient classical and quantum computing resources.

\section*{Acknowledgements}

This work has been supported by the Academy of Finland (project DEQSE 349945) and Business Finland (projects TORQS 8582/31/2022, EM4QS 155/31/2024 and SeQuSoS 112/31/2024).\\
\\
GitHub repository for the \texttt{q8s-kernel} project: https://github.com/torqs-project/q8s-kernel

\end{document}